\documentclass[twocolumn,english,showpacs,showkeys,superscriptaddress,citeautoscript]{revtex4-2}
\usepackage[T1]{fontenc}
\usepackage[latin9]{inputenc}
\usepackage[active]{srcltx}
\usepackage{amsmath}
\usepackage{amssymb}
\usepackage{graphicx}
\usepackage{esint}

\makeatletter

\providecommand{\tabularnewline}{\\}

\usepackage{babel}
\usepackage{rotating}

\newcommand{\updwn}{\rotatebox[origin=c]{90}{$\leftrightharpoons$}}
\DeclareMathOperator*{\Motimes}{\text{\raisebox{0.25ex}{\scalebox{0.6}{$\bigotimes$}}}}

\makeatother

\usepackage{babel}
\begin{document}
\title{Unusual low-temperature behavior in the half-filled band of the one-dimensional
extended Hubbard model in atomic limit}
\author{Onofre Rojas}
\affiliation{Departamento de Física, Universidade Federal de Lavras, CP 3037, 37200-900,
Lavras-MG, Brazil}
\author{S. M. de Souza}
\affiliation{Departamento de Física, Universidade Federal de Lavras, CP 3037, 37200-900,
Lavras-MG, Brazil}
\author{J. Torrico}
\affiliation{Departamento de Física, Instituto de Cíencias Exatas, Universidade
Federal de Alfenas, 37133-840 Alfenas, Minas Gerais, Brazil}
\author{L. M. Verissimo}
\affiliation{Donostia International Physics Center, Paseo Manuel de Lardizabal
4, E-20018 San Sebastián, Spain}
\affiliation{Instituto de Física, Universidade Federal de Alagoas, 57072-970 Maceió,
Alagoas, Brazil}
\author{M. S. S. Pereira}
\affiliation{Instituto de Física, Universidade Federal de Alagoas, 57072-970 Maceió,
Alagoas, Brazil}
\author{M. L. Lyra}
\affiliation{Instituto de Física, Universidade Federal de Alagoas, 57072-970 Maceió,
Alagoas, Brazil}
\author{Oleg Derzhko}
\affiliation{Institute for Condensed Matter Physics, National Academy of Sciences
of Ukraine, Svientsitskii Street 1, 79011 L'viv, Ukraine}
\begin{abstract}
Recently, a kind of finite-temperature pseudo-transition was observed
in several quasi-one-dimensional models. In this work, we consider
a genuine one-dimensional extended Hubbard model in the atomic limit,
influenced by an external magnetic field and with the arbitrary number
of particles controlled by the chemical potential. The one-dimensional
extended Hubbard model in the atomic limit was initially studied in
the seventies and has been investigated over the past decades, but
it still surprises us today with its fascinating properties. We rigorously
analyze its low-temperature behavior using the transfer matrix technique
and provide accurate numerical results. Our analysis confirms that
there is an anomalous behavior in the half-filled band, specifically
occurring between the alternating pair ($AP$) and paramagnetic ($PM$)
phases at zero temperature. Previous investigations did not deeply
identify this anomalous behavior, maybe due to the numerical simplicity
of the model, but from analytical point of view this is not so easy
to manipulate algebraically because one needs to solve an algebraic
cubic equation. In this study, we explore this behavior and clearly
distinguish the pseudo-transition, which could easily be mistaken
with a real phase transition. This anomalous behavior mimics features
of both first- and second-order phase transitions. However, due to
its nature, we cannot expect a finite-temperature phase transition
in this model. 
\end{abstract}
\date{\today}
\maketitle

\section{Introduction}

Recent studies on various effective-one-dimensional models with short-range
interactions have revealed intriguing thermal behaviors, resembling
first and second-order phase transitions \citep{pseudo}. This peculiar
behavior to be called further pseudo-transition \cite{pseudo} is
also dubbed ultranarrow phase crossover \cite{W-Yin-1,W-Yin-2}, thermal
pseudo-transition \cite{jozef24}, or curious thermodynamics that
resembles a phase transition \cite{chapman}. Pseudo-transitions have
been further analyzed in reference \cite{Isaac}, focusing on spin
correlation functions. The simplest models where this unusual behavior
arises are in the Ising diamond chain \citep{Strecka-ising,W-Yin-2,W-Yin-3}.
Another similar decorated Ising chain discussed is the Ising sawtooth-like
chain model \citep{W-Yin-3}, along with two and three-leg Ising ladder
models \citep{hutak21,W-Yin-1,W-Yin-4}. Pseudo-transitions also appear
in other models, such as the Ising-Heisenberg diamond chain \citep{torrico,torrico2},
and the one-dimensional double-tetrahedral model with alternating
Ising spins and delocalized electrons \citep{Galisova}. Similar phenomena
are observed in ladder models with Ising-Heisenberg coupling \citep{on-strk}
and triangular tube models \citep{strk-cav}, highlighting a pattern
of pseudo-transition. Further investigations \citep{hutak21,unv-cr-exp,krokhmalskii21}
have shown some kind of universality of power-law exponents, while
still satisfying the Rushbrooke inequality. All the above models are
related to \textit{classical} and \textit{classical}-quantum spin
models; here by \textit{classical} we mean that the Hamiltonian does
not contain non-commuting terms. However, models of other natures,
like the extended Hubbard model in a diamond chain structure \citep{psd-hub-dmd},
and the Potts model on a diamond chain structure, have also exhibited
this unusual behavior. On the other hand, exploring pseudo-transitions
in genuine one-dimensional systems without decoration couplings is
rather interesting. In this sense, the Zimm-Bragg-Potts model was
recently explored and found to exhibit this anomalous behavior \citep{Potts-psd}.

On the other hand, the Hubbard model \citep{Hubbard} stands as a
foundational model in the modern theory of strongly correlated electrons,
playing a paradigmatic role in the study of electronic correlations
in quantum materials. This model is particularly significant in contexts
where interactions are crucial. The Bethe ansatz method helps to understand
this model better, especially in figuring out how electrons behave
in different conditions and how they interact with each other \citep{Essler}.
The one-dimensional extended Hubbard model is a simplified theoretical
model that describes the behavior of electrons in a one-dimensional
chain, which additionally includes nearest-neighbor interaction energy
terms. The nearest-neighbor interaction energy term describes the
Coulomb repulsion between electrons occupying neighboring sites. The
one-dimensional extended Hubbard model has been widely studied in
the literature due to its relevance in understanding the electronic
properties of one-dimensional materials such as carbon nanotubes and
organic conductors \citep{Wang-09}. It also serves as a simple prototype
model for studying strongly correlated electron systems \citep{Dagotto}.
On the other hand, several theoretical investigations have focused
on the extended Hubbard model in the one-dimensional case. Numerous
investigations regarding the half-filled ground states of the extended
Hubbard model have been conducted. Tsuchiizu and Furusaki \citep{Tsuchiizu}
revisited the ground-state phase diagram of the one-dimensional half-filled
extended Hubbard model, revealing different phases and transitions
using a renormalization-group approach. Glocke \textit{et al.} \citep{Glocke}
utilized density-matrix renormalization group methods on transfer
matrices to study the thermodynamics of the one-dimensional extended
Hubbard model at half-filling. This highlights the detection of phase
transitions through standard thermodynamic measures like isothermal
compressibility and magnetic susceptibility. The study identifies
a unique phase with long-range dimer order and delineates the phase
diagram, comparing it with quantum Monte Carlo studies \citep{Sengupta,Sandvik}.
The phase diagram and power-law exponents of the one-dimensional $U-V$
model at quarter-filling were identified using exact diagonalization
and various limit results, identifying a transition from Luttinger
liquid to charge density wave insulator, noting dominant superconducting
or spin density wave fluctuations depending on the value of $V$ \citep{Zotos}.
Further investigation was also performed in reference \citep{zhang}.

Recent experiments have successfully demonstrated the realization
of an extended fermionic Hubbard model using a two-dimensional lattice
composed of dopant-based quantum dots. Quantum dots, often likened
to artificial atoms, can be accurately arranged into structures resembling
artificial molecules and lattices. These arrangements offer adjustable
hopping amplitudes and interaction strengths, as well as the ability
to design specific point symmetries. This advancement marks a significant
step in the exploration of complex quantum systems and could pave
the way for new insights into the behaviors of electronic systems
\citep{wang}. Recent angle-resolved photoemission spectroscopy (ARPES)
studies on the one-dimensional extended Hubbard model, employing bosonization
and time-dependent calculations, reveal insights into electron-phonon
coupling and interactions in one-dimensional systems \citep{Hx-Wang}.
Earlier, Epstein \textit{et al.} \citep{Epstein} discussed the metal-insulator
transition of N-methyl phenazinium (NMP) tetracyanoquinodimethane
(TCNQ) based on the strongly correlated Hubbard model ($t\ll U$).
A considerable number of theoretical studies of this model have been
undertaken, and many of its properties are now well known. However,
this simple model still surprises us with unexpected features, which
are the focus in the following sections.

Although the simplest version of the extended Hubbard model has been
considerably studied and applied to various physical systems, here
we consider a typical one-dimensional extended Hubbard model in the
atomic limit (neglecting the hopping term), 
\begin{alignat}{1}
\boldsymbol{H}= & \sum_{i=1}^{N}\left[U\boldsymbol{n}_{i,\uparrow}\boldsymbol{n}_{i,\downarrow}+V\boldsymbol{n}_{i}\boldsymbol{n}_{i+1}\right.\nonumber \\
 & \left.-\mu(\boldsymbol{n}_{i,\uparrow}+\boldsymbol{n}_{i,\downarrow})-h(\boldsymbol{n}_{i,\uparrow}-\boldsymbol{n}_{i,\downarrow})\right],
\end{alignat}
where $U$ is the on-site Coulomb interaction, $V$ is the Coulomb
interaction between electrons on the neighboring sites, $\mu$ is
the chemical potential, and $\boldsymbol{n}_{i,\sigma}$ is the corresponding
number operator at site $i$, with spin $\sigma=\{\uparrow,\downarrow\}$
and $\boldsymbol{n}_{i}=\boldsymbol{n}_{i,\uparrow}+\boldsymbol{n}_{i,\downarrow}$.
And the last term reports the contribution of external magnetic field
$h$. Despite its simplicity, the present model still astonishes us
by providing further interesting anomalous properties not previously
elucidated elsewhere, through careful analysis.

The present work is organized as follows: In Sec. 2, we give the thermodynamics
of the one-dimensional extended Hubbard model in the atomic limit,
identifying each eigenvalue of the cubic root and determining which
one is the largest, a topic not previously elucidated. In Sec. 3,
we analyze a peculiar property in the low-temperature region and explore
the anomalous behavior, along with the corresponding region where
the main properties undergo significant changes at this anomalous
temperature. This phenomenon is what we refer to as the quasi-phase
diagram \cite{pseudo}. In Sec. 4, we explore additional physical
quantities, reporting the influence of pseudo-transition. Finally,
in Sec. 5, we present our conclusions.

\section{Thermodynamics of the Model}

In the 1970s, the thermodynamics of the one-dimensional extended Hubbard
model in the atomic limit garnered significant interest, with pioneering
analyses by Bari \citep{bari}, Beni and Pincus \citep{beni-pincus},
and Gallinar \citep{gallinar} employing the transfer matrix approach.
These studies examined specific heat, static magnetic susceptibility,
and density-density correlation functions at various temperatures,
particularly focusing on half-filled band and briefly on the quarter-filled
band case with infinite intra-atomic Coulomb repulsion. Later, Mancini
and Mancini \citep{mancini05,Mancini} advanced this research using
Green's function and equations of motion formalism, finding four distinct
phases and diverse charge orderings at zero temperature. Their work
also considered the influence of external magnetic fields \citep{mancini09}
on thermodynamic properties like magnetization and specific heat,
identifying critical fields associated with polarization levels.

\subsection{Transfer matrix\protect\label{subsec:Transfer-matrix}}

In order to express the transfer matrix of the model, we use the following
natural basis $\{|0\rangle,|\uparrow\rangle,|\downarrow\rangle,|\updwn\rangle\}$.
The first state corresponds to the vacuum state, the second state
denotes the spin-up state, the third state is the spin-down state,
and the fourth state corresponds to two spins with opposite spins
on same site.

In principle, this model can be solved using the transfer matrix technique
\citep{beni-pincus,baxter-book}, and the transfer matrix is given
by 
\begin{equation}
\mathbf{W}=\left[\begin{array}{cccc}
1 & yw_{0,1} & y^{-1}w_{0,1} & w_{0,2}\\
yw_{0,1} & y^{2}w_{1,1} & w_{1,1} & yw_{1,2}\\
y^{-1}w_{0,1} & w_{1,1} & y^{-2}w_{1,1} & y^{-1}w_{1,2}\\
w_{0,2} & yw_{1,2} & y^{-1}w_{1,2} & w_{2,2}
\end{array}\right],\label{eq:V}
\end{equation}
where $w_{0,1}={\rm e}^{\beta\mu/2}$, $w_{0,2}={\rm e}^{\beta(\mu-U/2)}$,
$w_{1,1}={\rm e}^{\beta(\mu-V)}$, $w_{1,2}={\rm e}^{\beta(3\mu/2-2V-U/2)}$,
$w_{2,2}={\rm e}^{\beta(2\mu-4V-U)}$, $y={\rm e}^{\beta h/2}$, and
$\beta$ is the inverse temperature, $\beta=1/(k_{B}T)$.

Certainly, we can proceed by calculating $\text{det}(\mathbf{W}-\lambda\mathbf{1})=0$
to obtain the eigenvalues of the transfer matrix, as was previously
done by Beni and Pincus \cite{beni-pincus}. Alternatively, we can
manipulate the transfer matrix for convenience, considering the spin
inversion symmetry, in order to simplify our perturbative approach
calculations later on. Thus, we can employ a new set of basis vectors
$\{|0\rangle,|\updownarrow\rangle,|\updwn\rangle,|\leftrightarrow\rangle\}$,
where $|\updownarrow\rangle=\left(y|\uparrow\rangle+y^{-1}|\downarrow\rangle\right)/\sqrt{z}$
corresponds to the symmetric state and $|\leftrightarrow\rangle=\left(y|\downarrow\rangle-y^{-1}|\uparrow\rangle\right)/\sqrt{z}$
denotes the antisymmetric state, with $z=y^{2}+y^{-2}=2\cosh(\beta h)$.
Note that these states are invariant under simultaneous spin inversion
and magnetic field inversion. In this new basis, the transfer matrix
\eqref{eq:V} simply becomes:  
\begin{equation}
\mathbf{W}=\left[\begin{array}{cccc}
1 & w_{0,1}\sqrt{z} & w_{0,2} & 0\\
w_{0,1}\sqrt{z} & w_{1,1}z & w_{1,2}\sqrt{z} & 0\\
w_{0,2} & w_{1,2}\sqrt{z} & w_{2,2} & 0\\
0 & 0 & 0 & 0
\end{array}\right]=\left[\begin{array}{cc}
\mathbf{V} & \mathbf{0}\\
\mathbf{0} & 0
\end{array}\right],\label{eq:V-1}
\end{equation}
where the matrix $\mathbf{V}$, expressed in the basis $\{|0\rangle,|\updownarrow\rangle,|\updwn\rangle\}$,
results in  
\begin{equation}
\mathbf{V}=\left[\begin{array}{ccc}
1 & w_{0,1}\sqrt{z} & w_{0,2}\\
w_{0,1}\sqrt{z} & w_{1,1}z & w_{1,2}\sqrt{z}\\
w_{0,2} & w_{1,2}\sqrt{z} & w_{2,2}
\end{array}\right].
\end{equation}
Obviously, the eigenvalue corresponding to the state $|\leftrightarrow\rangle$
is null.

It is straightforward to diagonalize the transfer matrix by solving
the determinant equation ${\rm det}(\mathbf{V}-\lambda\mathbf{1})=0$,
which leads to the following secular equation 
\begin{equation}
\lambda^{3}+a_{2}\lambda^{2}+a_{1}\lambda+a_{0}=0,\label{eq:cub-eq}
\end{equation}
where the coefficients result in 
\begin{alignat}{1}
a_{0}= & z\left(w_{1,2}^{2}+w_{0,2}^{2}w_{1,1}+w_{0,1}^{2}w_{2,2}\right)\nonumber \\
 & -z\left(w_{1,1}w_{2,2}-2w_{0,2}w_{0,1}w_{1,2}\right),\nonumber \\
a_{1}= & z\left(w_{1,1}+w_{1,1}w_{2,2}-w_{0,1}^{2}-w_{1,2}^{2}\right)+w_{2,2}-w_{0,2}^{2},\nonumber \\
a_{2}= & -1-zw_{1,1}-w_{2,2}.\label{eq:coefs}
\end{alignat}
Therefore, the roots of the algebraic cubic equation \eqref{eq:cub-eq}
can be expressed conveniently using trigonometric function, i.e.,
\begin{equation}
\lambda_{j}=2\sqrt{Q}\cos\left(\tfrac{\phi-2\pi j}{3}\right)-\frac{1}{3}a_{2},\quad j=0,1,2,\label{eq:sol-cub}
\end{equation}
with 
\begin{alignat}{1}
\phi & =\arccos\left(\tfrac{R}{\sqrt{Q^{3}}}\right),\\
Q & ={\displaystyle \frac{a_{2}^{2}-3a_{1}}{9},}\label{eq:Q=000020real}\\
R & ={\displaystyle \frac{9a_{1}a_{2}-27a_{0}-2a_{2}^{3}}{54}.}\label{eq:=000020R=000020real}
\end{alignat}
Note that, it is sufficient to restrict the cubic equation solution
without loss of generality to the interval of $0<\phi<\pi$; thus
the eigenvalues are ordered as follows: $\lambda_{0}>\lambda_{1}>\lambda_{2}$.
This criterion is discussed in more detail in reference \citep{psd-hub-dmd}.
To analyze the characteristics of the eigenvalues in the interval
$0<\phi<\pi$, after some trigonometric manipulation, we have 
\begin{alignat}{1}
\sqrt{Q}-\tfrac{a_{2}}{3}<\lambda_{0}<2\sqrt{Q}-\tfrac{a_{2}}{3},
\end{alignat}
obviously $\lambda_{0}$ is definitely positive, because $a_{2}<0$.
Similarly, for the second eigenvalue, we can express 
\begin{alignat}{1}
-\sqrt{Q}-\tfrac{a_{2}}{3}<\lambda_{1}<\sqrt{Q}-\tfrac{a_{2}}{3},
\end{alignat}
in this case, $\lambda_{1}$ can be positive or negative depending
on the Hamiltonian parameters. For the last eigenvalue, the corresponding
interval can be expressed as 
\begin{alignat}{1}
-2\sqrt{Q}-\tfrac{a_{2}}{3}<\lambda_{2}<-\sqrt{Q}-\tfrac{a_{2}}{3},
\end{alignat}
similarly, $\lambda_{2}$ can be positive or negative. Particularly
to satisfy $\lambda_{2}<0$, the following condition must be met:
$\sqrt{Q}>-\tfrac{a_{2}}{3}$, which implies that $a_{1}<0$. It is
noteworthy that for other intervals, the arrangement of the cubic
root solutions merely exchanges; for details, see the Table 2 of reference
\citep{psd-hub-dmd}.

As said above, the fourth eigenvalue $\lambda_{3}$ of the transfer
matrix \eqref{eq:V} becomes null.

\subsection{Thermodynamic quantities and correlators \protect\label{subsec:Thermo}}

To analyze thermodynamic quantities, we use the grand partition function
for a chain consisting of $N$ site: 
\begin{eqnarray}
\Xi(T,\mu,h,N)=\lambda_{0}^{N}+\lambda_{1}^{N}+\lambda_{2}^{N}.\label{part-func}
\end{eqnarray}
Given the hierarchy of the eigenvalues ($\lambda_{0}>\lambda_{1}>\lambda_{2}$),
one can determine the grand potential per site in the thermodynamic
limit $N\to\infty$, which is solely dictated by the largest eigenvalue
of the transfer matrix: 
\begin{alignat}{1}
\Omega(T,\mu,h)= & -k_{B}T\lim_{N\rightarrow\infty}\frac{\ln\left[\Xi(T,\mu,h,N)\right]}{N}\nonumber \\
= & -k_{B}T\ln\lambda_{0}.\label{grand-pot}
\end{alignat}
Here, we introduce several useful thermodynamic quantities; all these
quantities will be quantified per site. The entropy is calculated
as ${\cal S}=-\frac{\partial\Omega}{\partial T}$; the enthalpy is
given by $\mathcal{E}=-\frac{\partial\ln\left(\lambda_{0}\right)}{\partial\beta}=k_{B}T^{2}\frac{\partial\ln\left(\lambda_{0}\right)}{\partial T}$;
analogously we can obtain specific heat at constant chemical potential
$C=T\frac{\partial{\cal S}}{\partial T}=\frac{\partial{\cal E}}{\partial T}$
\cite{v-derzhk}; the magnetization can be expressed as $M=-\frac{\partial\Omega}{\partial h}$;
magnetic susceptibility $\chi=\frac{\partial M}{\partial h}$; the
electron density $\rho$ is determined by $\rho=-\frac{\partial\Omega}{\partial\mu}$;
and the isothermal compressibility is derived from $\kappa=\tfrac{1}{\rho^{2}}\frac{\partial\rho}{\partial\mu}$.

It is also feasible to determine other quantities using $\rho$ and
$M$, which can be expressed as $\rho=\langle\boldsymbol{n}_{\uparrow}\rangle+\langle\boldsymbol{n}_{\downarrow}\rangle$
and $M=\langle\boldsymbol{n}_{\uparrow}\rangle-\langle\boldsymbol{n}_{\downarrow}\rangle$,
leading to the expressions 
\begin{alignat}{2}
\langle\boldsymbol{n}_{\uparrow}\rangle= & \frac{\rho+M}{2}\quad\text{and}\quad & \langle\boldsymbol{n}_{\downarrow}\rangle= & \frac{\rho-M}{2}.
\end{alignat}
Similarly, we can also obtain the following quantities 
\begin{alignat}{2}
\langle\boldsymbol{n}_{\uparrow}\boldsymbol{n}_{\downarrow}\rangle= & \frac{\partial\Omega}{\partial U},\quad & \langle\boldsymbol{n}_{i}\boldsymbol{n}_{i+1}\rangle= & \frac{\partial\Omega}{\partial V},
\end{alignat}
where we define 
\begin{equation}
\langle\boldsymbol{n}_{i}\boldsymbol{n}_{i+1}\rangle\equiv\langle\left(\boldsymbol{n}_{i,\uparrow}+\boldsymbol{n}_{i,\downarrow}\right)\left(\boldsymbol{n}_{i+1,\uparrow}+\boldsymbol{n}_{i+1,\downarrow}\right)\rangle.\label{eq:<n1n2>}
\end{equation}

Furthermore, for a transfer matrix with a non-degenerate and positively
defined spectrum, one can easily define the correlation length, $\xi$,
as the inverse logarithm of the ratio of the largest and the second-largest
eigenvalue. Although our transfer matrix eigenvalues spectrum is non-degenerate,
some eigenvalues could become negative. Therefore, we need to compare
the magnitude of each eigenvalue. For instance, focusing near the
pseudo-transition, the eigenvalues become $\lambda_{0}>0$ and, similarly,
$\lambda_{1}>0$ is also positive. Nevertheless, $\lambda_{2}<0$.
However, in terms of magnitude, we cannot determine whether $|\lambda_{1}|$
or $|\lambda_{2}|$ is the second largest. Depending on the Hamiltonian
parameters, we may have $|\lambda_{1}|>|\lambda_{2}|$ or $|\lambda_{1}|<|\lambda_{2}|$.
Therefore, for our case, we define the correlation length \cite{Alves-jr,Wang-Guo}
as follows: 
\begin{equation}
\xi=\left[\ln\left(\frac{\lambda_{0}}{\max(|\lambda_{1}|,|\lambda_{2}|)}\right)\right]^{-1}.\label{xi}
\end{equation}
Note that $\max(|\lambda_{1}|,|\lambda_{2}|)$ has nothing to do with
exchanging cubic root solutions but depends solely on the Hamiltonian
parameters.

Evidently the correlation length, $\xi$, becomes more intricate due
to the competition between the magnitudes of the second-largest eigenvalues.
This complexity in $\xi$ is more pronounced near the point where
the magnitudes of these eigenvalues compete. Far from this point of
competition, the behavior of $\xi$ could be more straightforwardly
described.

\section{Phase diagrams}

In what follows, we discuss the ground-state energy, focusing on peculiar
regions where anomalous behavior appears. Then we consider the low-temperature
case to discuss quasi-phase diagrams and define a peculiar temperature.

\subsection{Zero-temperature phase diagram}

In the absence of magnetic field $h$, the first ground state to consider
is the frustrated phase $FR_{1}$ 
\begin{alignat}{1}
|FR_{1}\rangle= & \Motimes_{j=1}^{N/2}|0,\sigma_{2j}\rangle\quad\text{or}\quad\Motimes_{j=1}^{N/2}|\sigma_{2j},0\rangle
\end{alignat}
with the respective ground-state energy 
\begin{alignat}{1}
E_{FR_{1}}= & -\frac{h}{2}-\frac{\mu}{2}.
\end{alignat}
Note that the frustrated ground-state energy becomes just at $h=0$.
For a non-null magnetic field, the state aligns with the magnetic
field, thereby becoming non-frustrated. The electron density for this
phase is given by $\rho=1/2$, commonly known as quarter-filled band.
The residual entropy (per site) of this phase is expressed as $\mathcal{S}=k_{B}\ln(2)/2$.

Another phase is the frustrated phase $FR_{2}$ given by 
\begin{alignat}{1}
|FR_{2}\rangle= & \Motimes_{j=1}^{N}|\updownarrow\rangle,\nonumber \\
= & \tfrac{1}{\sqrt{z}^{N}}\Motimes_{j=1}^{N}\left(y|\uparrow\rangle+y^{-1}|\downarrow\rangle\right).
\end{alignat}
The ground-state energy for this phase is 
\begin{alignat}{1}
E_{FR_{2}}= & V-h-\mu.
\end{alignat}
Here again, the frustrated ground state energy becomes only when $h=0$.
However, when $h\ne0$, the state gradually aligns with the magnetic
field, losing its frustration. The residual entropy of this phase
is $\mathcal{S}=k_{B}\ln(2)$, and the particle density is $\rho=1$
(half occupancy).

At a half-filling band or electron density $\rho=1$, we observe the
alternation pair phase $AP$ 
\begin{alignat}{1}
|AP\rangle= & \Motimes_{j=1}^{N/2}|0,\updwn\rangle\quad\text{or}\quad\Motimes_{j=1}^{N/2}|\updwn,0\rangle.
\end{alignat}
The corresponding ground-state energy is 
\begin{alignat}{1}
E_{AP}= & \frac{U}{2}-\mu.
\end{alignat}
Notably, the $AP$ phase has no residual entropy and can be identified
as charge density wave ($CDW$) \citep{Glocke}.

The third frustrated phase $FR_{3}$, with an electron density of
$\rho=3/2$, in the absence of a magnetic field, is described as 
\begin{alignat}{1}
|FR_{3}\rangle= & \Motimes_{j=1}^{N/2}|\updwn,\sigma_{2j}\rangle\quad\text{or}\quad\Motimes_{j=1}^{N/2}|\sigma_{2j},\updwn\rangle.
\end{alignat}
The ground-state energy for this phase is 
\begin{alignat}{1}
E_{FR_{3}}= & \frac{U}{2}+2V-\frac{h}{2}-\frac{3\mu}{2}.
\end{alignat}
Again the ground state is frustrated at $h=0$, whereas for $h\ne0$
the system loses its frustration. For $h=0$, the corresponding residual
entropy leads to $\mathcal{S}=k_{B}\ln(2)/2$.

Lastly, in the fully filled phase $FF$, the ground state results
in 
\begin{alignat}{1}
|FF\rangle= & \Motimes_{j=1}^{N}|\updwn\rangle.
\end{alignat}
The corresponding ground-state energy is 
\begin{alignat}{1}
E_{FF}= & U-4V-2\mu.
\end{alignat}
The electron density in this phase is $\rho=2$, and there is no residual
entropy.

It is important to note that in all the aforementioned states, the
magnetization obviously becomes $M=0$ at $h=0$. 
Furthermore, including a magnetic field $h$ into the phase diagrams
generally results in properties similar to those observed in the absence
of a magnetic field (not illustrated). The main distinction is that
the regions previously identified as frustrated ($FR$) become unfrustrated
in the presence of a magnetic field. An equivalent analysis was earlier
performed by Mancini and Mancini \citep{Mancini,mancini09}, offering
a different perspective on the phase diagram. For example, in the
$FR_{1}$ (quarter filled) and $FR_{3}$ (three quarter filled) regions,
under a sufficiently strong magnetic field, the spin arrangements
align parallel to the external magnetic field. In contrast, the $FR_{2}$
phase, characterized by one particle per site with randomly oriented
spins, begins to show alignment under the influence of a magnetic
field. As the magnetic field strength increases, the spins start aligning
with the external field, transitioning into a fully polarized or paramagnetic
phase $PM$.

\begin{figure}
\includegraphics[scale=0.54]{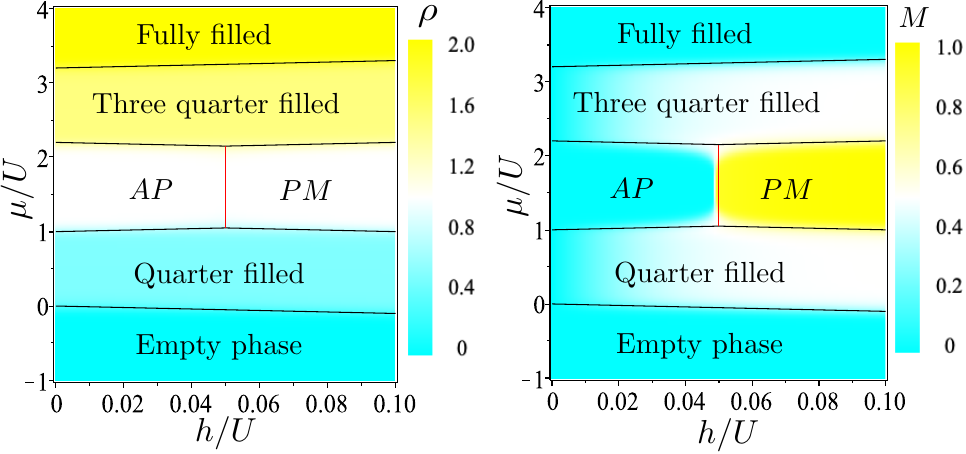} \caption{\protect\label{fig:ph-h-mu} Phase diagram in the $\mu/U-h/U$ plane,
with solid lines, describes the zero-temperature phase diagram for
the fixed $V=0.55$. Meanwhile, the background color density plot
corresponds to electron density (left) and magnetization (right),
assuming the low temperature $T=0.03$.}
\end{figure}

For a clearer illustration, we present in Fig. \ref{fig:ph-h-mu}
the phase diagram in the $h/U-\mu/U$ plane for fixed $V=0.55$. Black
solid lines describe the zero-temperature phase diagram, while the
red solid line delineates an unusual phase boundary where anomalous
behavior emerges. In the same plot, we also incorporate a color density
plot corresponding to electron density $\rho$ (left) and magnetization
$M$ (right) in the low temperature region $T=0.03$. In the left
panel, we observe a phase transition between alternating pair ($AP$)
and paramagnetic ($PM$) phases in the half-filled region, where a
constant density is clearly visible. In contrast, we note a distinct
change in magnetization from $M=0$ in the $AP$ phase to $M=1$ in
the $PM$ phase. The half-filling band region be our focus from now
on.

\begin{table}
\caption{\protect\label{tab1}The second column provides zero-temperature phase
boundary conditions, the third column gives the chemical potential
at the phase boundary, while the fourth column reports the associated
residual boundary entropy. Note that for the $FR_{2}-AP$ there is
no dependence on $\mu$ according to the second column and therefore
the corresponding row in the third column is empty.}
\begin{tabular}{|c|c|c|c|}
\hline 
Boundary  & $h$  & $\mu$  & $\mathcal{S}/k_{B}$\tabularnewline
\hline 
\hline 
$FF-FR_{3}$  & $\mu-U-2V$  & $h+U+2V$  & $\ln(2)$\tabularnewline
\hline 
$FR_{3}-AP$  & $4V-\mu$  & $4V-h$  & $\ln(3)/2$\tabularnewline
\hline 
$FR_{3}-FR_{2}$  & $\mu-U-4V$  & $h+U+4V$  & $\ln(1+\sqrt{3})$\tabularnewline
\hline 
$FR_{2}-AP$  & $V-\tfrac{U}{2}$  &  & $\ln(2)$\tabularnewline
\hline 
$FR_{1}-AP$  & $\mu-U$  & $\mu+U$  & $\ln(3)/2$\tabularnewline
\hline 
$FR_{1}-FR_{2}$  & $2V-\mu$  & $2V-h$  & $\ln(1+\sqrt{3})$\tabularnewline
\hline 
\end{tabular}
\end{table}

Table \ref{tab1} presents the behavior of the phase boundaries: The
second column lists the magnetic field $h$ at each phase boundary,
and the fourth column details the residual entropy at the different
boundaries. Typically, at the interface between two phases, the residual
entropy is higher than in the adjacent phases. However, in Table \ref{tab1},
the phase boundary between $FR_{2}$ and $AP$ exhibits an unusual
behavior, the residual boundary entropy does not exceed the residual
entropy of $FR_{2}$ \cite{BJP20}. This anomaly leads to unexpected
behavior at finite temperatures, as discussed in Sec. \ref{subsec:Low-temperature-quasi-phase-diag}.

\begin{figure}
\includegraphics[scale=0.45]{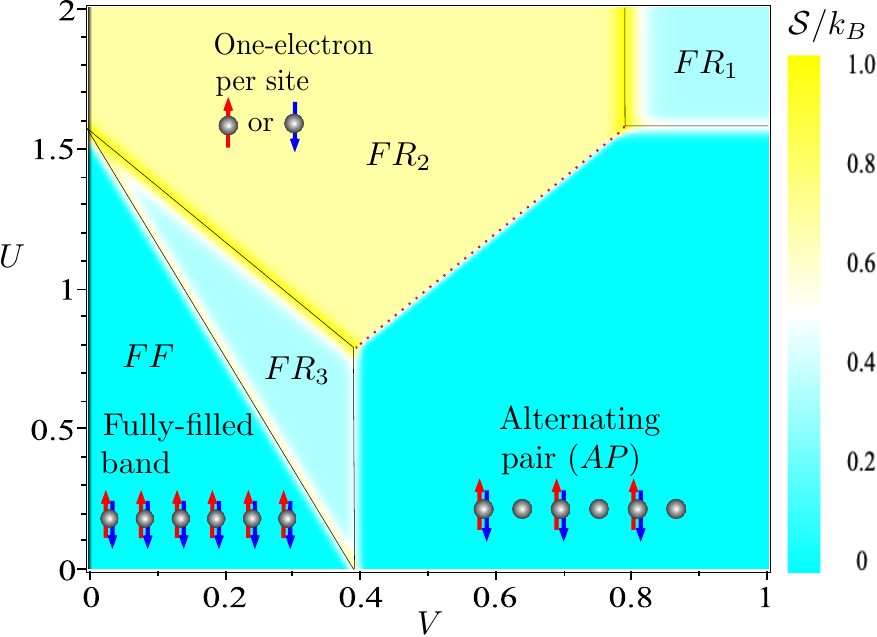}\caption{\protect\label{fig:Ph-DG-UV} Phase diagram in the $U-V$ plane under
the assumption of zero magnetic field and $\mu=1.58$. The solid line
represents the phase diagram at zero temperature, while the background
density plot illustrates the entropy at $T=0.01$.}
\end{figure}

The phase diagram in the $U-V$ plane, assuming a zero magnetic field
and considering a suitable chemical potential ($\mu=1.58$) to enhance
the anomalous behavior at a half-filled band, is illustrated in Fig.
\ref{fig:Ph-DG-UV}. This phase diagram features a fully filled band
phase $FF$, an alternating pair phase $AP$, and three frustrated
phases. The first frustrated phase $FR_{1}$ corresponds to a quarter-filled
band of electrons, with spins that can point either up or down, an
electron density of $\rho=1/2$, and a residual entropy of $\mathcal{S}=k_{B}\ln(2)/2$.
In the half-filled region, there is another frustrated phase $FR_{2}$
with one electron per site and spins that can also randomly point
up or down, having a residual entropy of $\mathcal{S}=k_{B}\ln(2)$.
The third frustrated phase $FR_{3}$ corresponds to a $3/4$-filled
band of electrons or a quarter-filled band of holes, with an electron
density of $\rho=3/2$ and a residual entropy of $\mathcal{S}=k_{B}\ln(2)/2$.
Additionally, the diagram shows a fully-filled band of electrons.
All these curves can be obtained from Table~\ref{tab1}, second column,
assuming $h=0$. Solid lines delineate standard phase boundaries,
while the dashed line indicates an anomalous boundary between two
half-filled regions.

\subsection{Anomalous behavior in low-temperature region and quasi-phase diagram
\protect\label{subsec:Low-temperature-quasi-phase-diag}}

In our exhaustive analysis, we explore a unique property emerging
in the low-temperature region, a phenomenon not extensively observed
or detailed in previous studies \citep{bari,beni-pincus,gallinar,Mancini,mancini09}.
Remarkably, our findings reveal intriguing phenomena within such a
simplistic model.

The background of Fig. \ref{fig:Ph-DG-UV} presents a density plot
of entropy in the $V-U$ plane at a low temperature ($T=0.01$). Our
focus is on the zero-temperature boundary between the $FR_{2}$ phase
and the $AP$ phase. Given the absence of true phase transitions,
with only crossover lines occurring at finite temperatures, we refer
to the zero-temperature phases at finite temperatures as quasi-phases.
In this low-temperature region, these phases are termed the quasi-frustrated
$qFR_{2}$ region and the quasi-alternating pair $qAP$ region or
quasi-$CDW$ region, where the arrangement of most spins remains similar
to their configurations at zero temperature.

\begin{figure}
\includegraphics[scale=0.5]{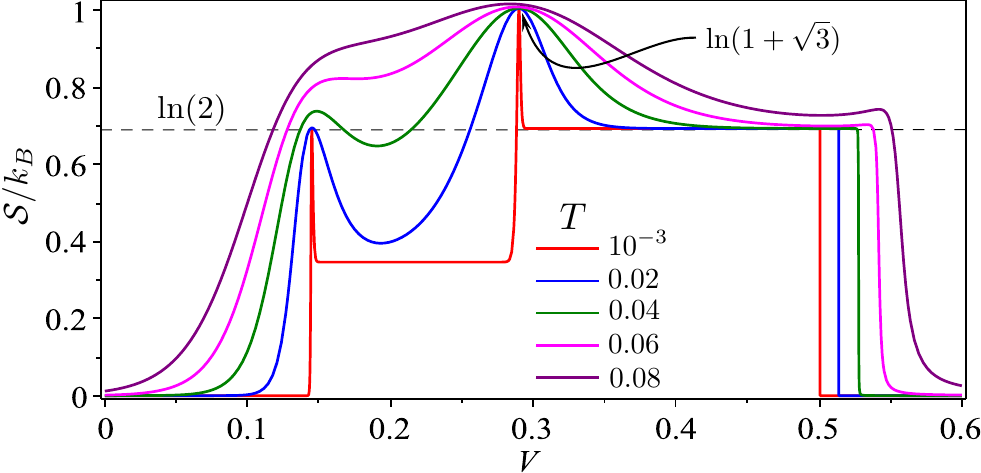}\caption{\protect\label{resd-S-V}Entropy as a function of $V$ for several
values of low temperatures, assuming fixed $U=1$ and $\mu=1.58$.}
\end{figure}

Another perspective of the background density plot illustrating the
entropy, as depicted in Fig. \ref{fig:Ph-DG-UV}, can be seen in Fig.
\ref{resd-S-V} for fixed values of $U=1$ and $\mu=1.58$, plotted
against $V$. Here, it is evident how residual entropy influences
thermal entropy. The peaks at $V\approx0.145$ and $V\approx0.29$
correspond to the standard interphases between $FF-FR_{3}$ (with
$\mathcal{S}=k_{B}\ln(2)$) and $FR_{3}-FR_{2}$ (with $\mathcal{S}=k_{B}\ln(1+\sqrt{3})$),
respectively. However, in the interphase between $FR_{2}-AP$ occurring
at $V=0.5$, no peaks are observed, indicating anomalous behavior.
Here, residual entropy leads to $\mathcal{S}=k_{B}\ln(2)$.

\begin{figure}
\includegraphics[scale=0.56]{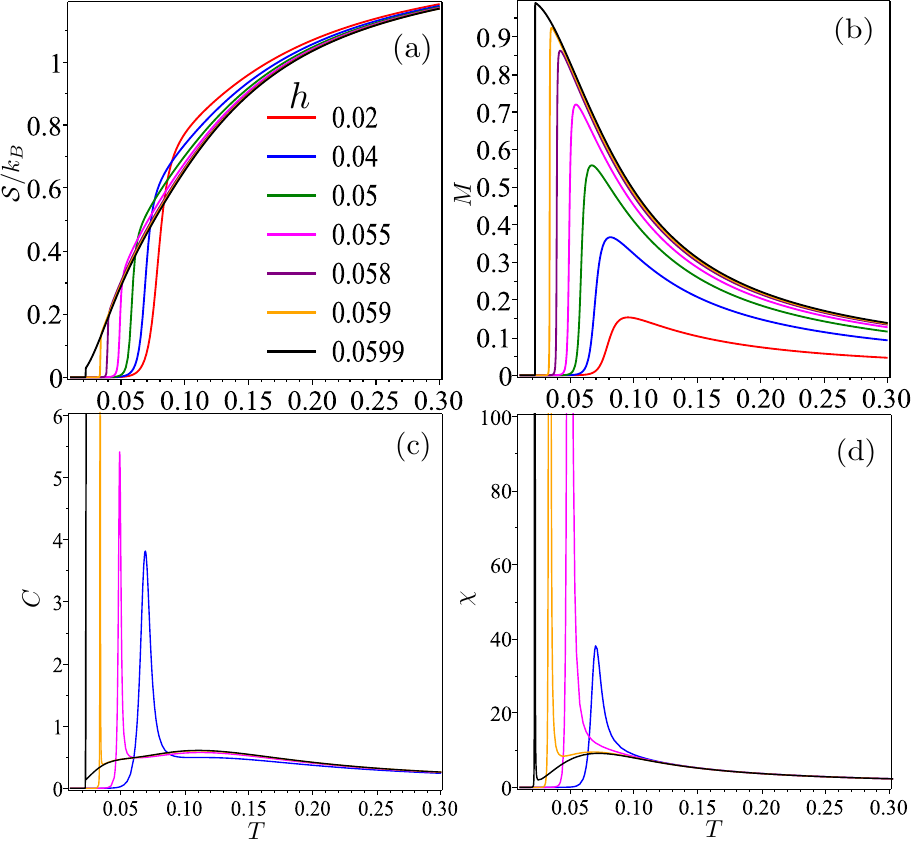} \caption{\protect\label{anm-qnts} (a) Entropy against $T$ for several values
of magnetic fields, assuming $U=1$ and $V=0.56$. (b) Magnetization
as a function of $T$. (c) Specific heat as a function of $T$. (d)
Magnetic susceptibility as a function of $T$. All curves correspond
to the legend given in (a).}
\end{figure}

Further evidence of this anomaly is explored in Fig. \ref{anm-qnts}.
Panel (a) presents the entropy (${\cal S}$) as a function of temperature
($T$) for several values of the external magnetic field, as indicated
within the panel. Here, one can clearly observe a swift change in
entropy at a specific low temperature region. Similarly, in panel
(b), we illustrate the magnetization ($M$) as a function of temperature
($T$); the colored curves refer to the same set of parameters as
in panel (a). Once again, we observe that the magnetization for all
sets of parameters is essentially null up to a certain temperature
where an unusual feature arises, followed by a rapid change in magnetization.
Almost full polarization is achieved above the anomalous temperature
with increasing magnetic fields. For moderately high temperatures,
standard thermal magnetization governs. Another quantity we examine
is the specific heat as a function of temperature under the same conditions
as in panel (a). Here, one can clearly observe very sharp peaks, with
the sharpness increasing at lower temperatures, akin to a second-order
phase transition, although it is merely a sharp peak with no divergence.
The magnetic susceptibility as a function of temperature is depicted
in panel (d), where we also observe a very sharp peak at a certain
temperature. Although it mimics a second-order phase transition, there
is no divergence, that would indicate a genuine phase transition.
Therefore, we will refer to these anomalous peaks as a ``false''
phase transition or pseudo-transition at a finite temperature.

\begin{figure}
\includegraphics[scale=0.57]{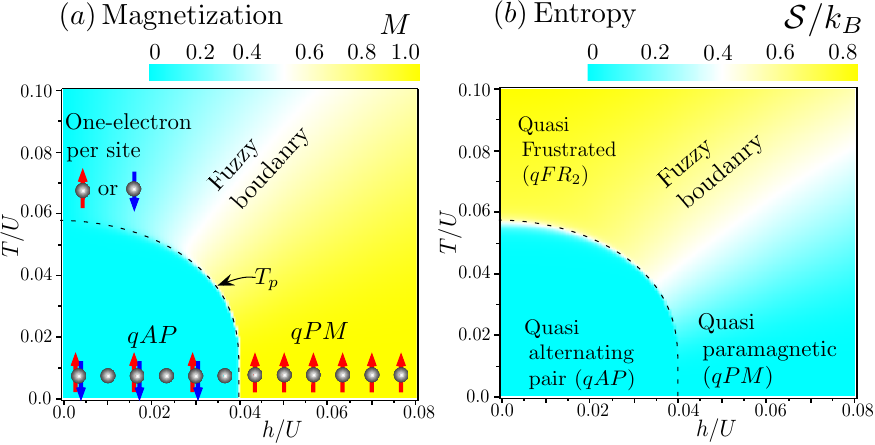} \caption{\protect\label{fig:Phase-diagram} Quasi-phase diagram depicting the
half-filled band in the $h/U-T/U$ plane, with fixed parameters $V/U=0.54$
and $\mu/U=1.58$. The dashed curve delineates the peculiar temperature
$T_{p}$. Panels (a) and (b) showcase density plots of magnetization
and entropy, respectively.}
\end{figure}

In exploring another intriguing aspect of the low-temperature region,
Fig. \ref{fig:Phase-diagram}a presents a density plot of magnetization
$M$ in the $h/U-T/U$ plane, assuming $V/U=0.54$ and $\mu/U=1.58$.
Here, the dashed line represents the peculiar temperature $T_{p}$.
Below a magnetic field of $h/U=0.04$ at sufficiently low temperatures,
the system resides in a quasi alternating-pair $qAP$ region, characterized
by particles predominantly arranged in alternating pairs. As the magnetic
field increases within this low temperature regime, the system goes
swiftly into a quasi-paramagnetic $qPM$ region, where most spins,
one per site, align parallel to the magnetic field. Conversely, at
very weak magnetic fields and increasing temperatures, the $FR_{2}$
frustrated system leads into a quasi-frustrated $qFR_{2}$ region.
The boundary between $qFR_{2}$ and $qPM$ is marked by a standard
crossover, or as visualized in the magnetization plot, a diffuse boundary,
appearing as a fuzzy region. In panel (a), one can observe the magnetization
shift between $qAP$ and $qPM$ regions, although the boundary between
$qAP$ and $qFR_{2}$ remains indistinct due to null magnetization
in both regions. Similarly, Fig. \ref{fig:Phase-diagram}b displays
the entropy $\mathcal{S}$, using the same parameters as panel (a).
Here, the crossover boundary, denoted by a dashed line, outlines the
peculiar temperature as a function of magnetic field. In contrast,
panel (b) shows that the boundary between $qAP$ and $qPM$ becomes
indistinguishable with almost null entropy, while the boundary between
$qAP$ and $qFR_{2}$ is sharply defined. This distinction is due
to the residual entropy of $FR_{2}$ being $\mathcal{S}=k_{B}\ln(2)$.
However, at temperatures above $T>0.06$, the entropy significantly
increases, becoming clearly larger than $\mathcal{S}\gtrsim0.7k_{B}$.

\subsection{Peculiar temperature condition \protect\label{subsec:Psd-crt}}

This detailed analysis further explores the interesting behavior in
the low-temperature region of the model. Previously, the peculiar
temperature was defined for models where the free energy involved
a simple square root expression \citep{pseudo,Isaac,unv-cr-exp}.
Defining the peculiar temperature generally poses a challenge, as
the only available indicator is an anomalous behavior in the low-temperature
region, marked by significant changes in entropy and magnetization,
along with sharp peaks in correlation length, specific heat, and magnetic
susceptibility. At very low temperatures, these peaks occur at roughly
the same temperature. However, as the anomalous behavior shifts to
higher temperatures $T_{p}$, these peaks appear at slightly different
temperatures for each quantity, complicating the definition of the
peculiar temperature.

An interesting observation is that this anomalous behavior is evident
in fundamental quantities such as the ratio between the second largest
and the largest eigenvalues. To analyze this more deeply, consider
the temperature and magnetic field dependence of the eigenvalues.
At a fixed magnetic field $h_{1}$, there exists a special temperature,
the peculiar temperature $T_{p}$, which satisfies the relation 
\begin{equation}
\left.\frac{\partial}{\partial T}\left(\tfrac{\lambda_{1}\left(T,h_{1}\right)}{\lambda_{0}\left(T,h_{1}\right)}\right)\right|_{T_{p}}=0.\label{eq:T_p}
\end{equation}
A similar analysis can be conducted by taking derivatives with respect
to the magnetic field of the ratio of the second largest to the largest
eigenvalues, assuming a fixed temperature $T_{p^{*}}$: 
\begin{equation}
\left.\frac{\partial}{\partial h}\left(\tfrac{\lambda_{1}\left(T_{p^{*}},h\right)}{\lambda_{0}\left(T_{p^{*}},h\right)}\right)\right|_{h_{1}}=0.\label{eq:T_p*}
\end{equation}
Comparing $T_{p^{*}}$ and $T_{p}$, we observe that this quantity
becomes slightly different as the temperature increases.

\begin{figure}
\includegraphics[scale=0.5]{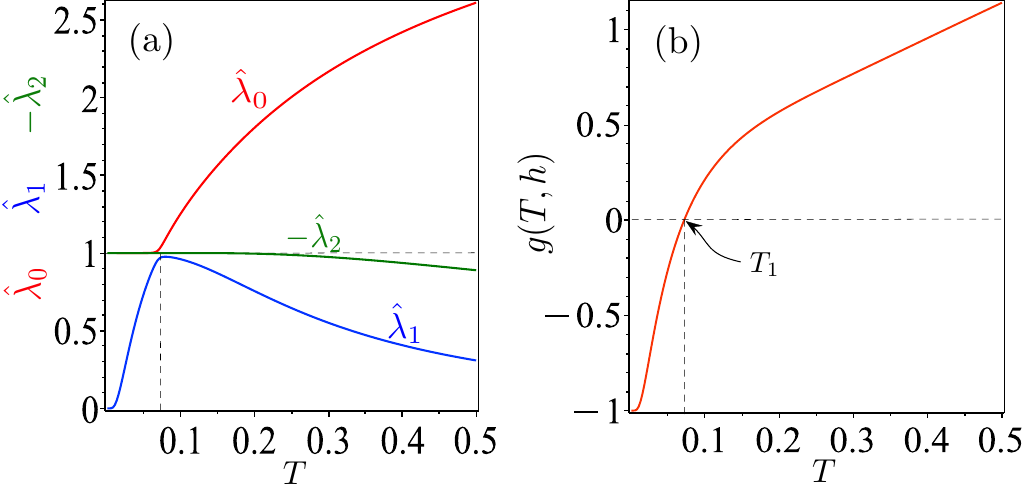} \caption{\protect\label{fig:Cubic-root-solution} (a) Cubic root solutions
$\hat{\lambda}_{j}$ as a function of temperature, under the condition
of a null magnetic field ($h=0$), with fixed values of chemical potential
($\mu=1.58$), on-site interaction strength ($U=1$), and nearest-neighbor
interaction strength ($V=0.55$). (b) Plot of $g(T,0)$ against temperature
$\ensuremath{T}$, using the same parameter set as in panel (a). }
\end{figure}

In order to explore further properties of the eigenvalues given in
\eqref{eq:sol-cub}, let us define $\hat{\lambda}_{j}(T,h)=\lambda_{j}{\rm e}^{\varepsilon_{0}(h)/k_{B}T}$,
where $\varepsilon_{0}(h)$ represents the ground-state energy. This
formulation, as discussed in reference \citep{pimenta}, primarily
aims to handle the ground-state energy explicitly, thereby circumventing
the issue of dealing with extremely large numbers. Thus, Fig. \ref{fig:Cubic-root-solution}a
depicts the three eigenvalues $\hat{\lambda}_{j}$ as functions of
temperature, under the conditions of a null magnetic field, $\mu=1.58$,
$U=1$, and $V=0.55$. Below the temperature $T_{1}$, $\hat{\lambda}_{2}$
approaches $-\hat{\lambda}_{0}$, whereas above $T_{1}$, $\hat{\lambda}_{2}$
approaches $-\hat{\lambda}_{1}$ but quickly diverges due to thermal
fluctuations.

Following this, we introduce the function 
\begin{equation}
g(T,h)=\hat{\lambda}_{0}(T,h)+\hat{\lambda}_{1}(T,h)+2\hat{\lambda}_{2}(T,h).\label{eq:g-func}
\end{equation}
This function, $g(T,h)$, can yield either positive or negative values.
Therefore, the condition 
\begin{equation}
g(T_{1},h_{1})=0,\label{eq:g-cond}
\end{equation}
for a fixed $h_{1}$ leads to the identification of a specific temperature,
$T_{1}$. This approach provides a practical method for examining
the temperature-dependent behavior of the system under study.

Panel (b) of the Fig. \ref{fig:Cubic-root-solution} illustrates the
function $g(T,h)$ as it varies with temperature, under the conditions
of a null magnetic field, $\mu=1.58$, $U=1$, and $V=0.55$. In this
representation, one can observe that the function $g(T,h)$ passes
through zero at a specific temperature, denoted as $T_{1}$. This
provides a clear visual indication of the peculiar point in the behavior
of the system.

Alternatively, the condition \eqref{eq:g-cond} can be simplified
through some algebraic manipulation, even in more general cases and
with generic coefficients of a cubic polynomial as outlined in \eqref{eq:cub-eq}.
This leads to a simple yet interesting result: 
\begin{equation}
2\tilde{a}_{2}^{3}+\tilde{a}_{1}\tilde{a}_{2}+\tilde{a}_{0}=0,\label{eq:T_g1}
\end{equation}
where $\tilde{a}_{j}$ is defined as $a_{j}(T_{1},h_{1})$.

Furthermore, we can use the expressions defined in \eqref{eq:coefs}
within the context provided by \eqref{eq:T_g1} to obtain after some
algebraic manipulation the following condition

\begin{equation}
z^{2}w_{1,1}^{2}+3zw_{1,1}-w_{0,2}^{2}+2=0.
\end{equation}
Given that $zw_{1,1}\sim w_{0,2}\gg1$ the dominant terms are $z^{2}w_{1,1}^{2}$
and $w_{0,2}^{2}$, while linear term $3zw_{1,1}$ and $2$ become
negligible and are thus omitted for the simplification. Consequently,
the above expression reduces simply to: 
\begin{equation}
w_{1,1}z-w_{0,2}=0.\label{eq:w-cnd-0}
\end{equation}

Furthermore, by employing a perturbative approach, as elaborated in
Appendix \ref{sec:Appdx=000020A} and illustrated through Eqs. \eqref{V011}
and \eqref{V002} for the unperturbed result, we reaffirm the condition
of the equation derived earlier, simplifying our findings into this
coherent expression.

In regions of low temperature, $T_{p}\approx T_{p^{*}}\approx T_{1}$,
which are very close to each other. When this occurs, we can observe
anomalous behavior, and we simply define it as the peculiar temperature,
$T_{p}$.

For the present model the condition \eqref{eq:w-cnd-0} provides a
sufficiently accurate condition for determining the peculiar temperature,
where $T_{p}\approx T_{1}$ holds. Specifically, the peculiar temperature
for a fixed magnetic field $h_{1}$ can be represented using the results
$w_{0,2}={\rm e}^{\beta(\mu-U/2)}$, $w_{1,1}={\rm e}^{\beta(\mu-V)}$,
and $z=2\cosh(\beta h)$, leading to the condition 
\begin{equation}
2\cosh(\beta_{1}h_{1})={\rm e}^{\beta_{1}(V-U/2)}.\label{eq:w-cond-h}
\end{equation}
Assuming $h_{1}>0$ and at very low-temperatures, we deduce that $h_{1}=V-\frac{U}{2}$.
Under these conditions, a pseudo-transition is observed at low temperatures.

Eventually, to refine our previous result, we can apply a first-order
perturbation approximation with respect to the non-half-filled contribution,
as detailed in Appendix \ref{sec:Appdx=000020A}. By using the results
given in Eqs. \eqref{eq:V1_11} and \eqref{eq:V1_02} we arrive at
the modified condition: 
\begin{equation}
w_{1,1}z-w_{0,2}+\frac{3}{2}(1+w_{2,2})=0.
\end{equation}
Moreover, further corrections can be applied up to a second-order
approximation, as detailed in the Appendix \ref{sec:Appdx=000020A}.

\begin{figure}
\includegraphics[scale=0.6]{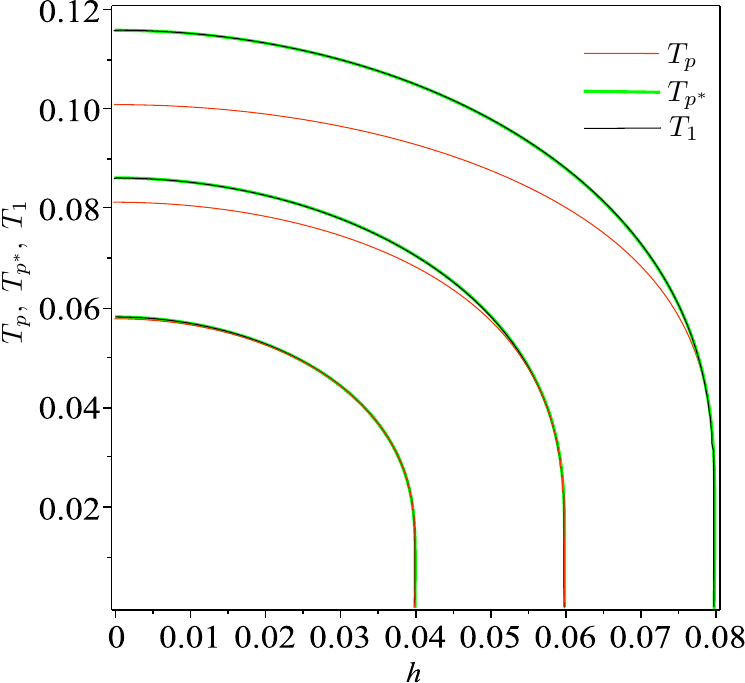} \caption{\protect\label{fig:Pseudo-crit} Plot of $T_{1}$ (thin black line),
$T_{p^{*}}$ (thick green line), and $T_{p}$ (thin red line) as functions
of $h$. The curves are obtained under the assumption of fixed parameters
$\mu=1.58$, $U=1$, and varying $V$ values: $V=0.54$ (inner curve),
$V=0.56$ (middle curve), and $V=0.58$ (outer curve).}
\end{figure}

In Fig. \ref{fig:Pseudo-crit}, the temperatures $T_{1}$ (represented
by a thin black line), $T_{p^{*}}$ (depicted by a thick green line),
and $T_{p}$ (illustrated with a thin red line) are shown as functions
of $h$, under fixed parameters $\mu=1.58$, $U=1$, and varying $V$
values: $0.54$ (inner curve), $0.56$ (middle curve), and $0.58$
(outer curve). For the inner curve, $T_{1}$, $T_{p^{*}}$, and $T_{p}$
are visually almost overlapping. However, as we observe the middle
curve, the discrepancy between $T_{1}\ensuremath{\approx}T_{p^{*}}$
and $T_{p}$ becomes more pronounced at higher temperatures. This
discrepancy is even more significant for the outer curve. Based on
this discrepancy, we can conclude that the peculiar transition is
marked for temperatures below $T_{p}\ensuremath{\lesssim}0.06$, where
all curves overlap (when $T_{1}\ensuremath{\approx}T_{p^{*}}$ and
$T_{p}$ start to diverge). A practical approach to observe the peculiar
temperature at higher temperatures is to consider when $\Delta T_{p}=T_{p}-T_{p^{*}}\rightarrow0$.
This is because when $\Delta T_{p}$ becomes significant, it becomes
less meaningful to refer to it as a peculiar temperature.

\begin{table}
\caption{\protect\label{tab:Psd-crt} Peculiar temperature for $\mu=1.58$,
$U=1$, and $V=0.56$. The second column is computed using equation
\eqref{eq:T_p}, the third column utilizes equation \eqref{eq:T_p*},
and the fourth column is determined by employing equation \eqref{eq:g-cond}.}
\begin{tabular}{|c|c|c|c|}
\hline 
$h$  & $T_{p}$  & $T_{p^{*}}$  & $T_{1}$\tabularnewline
\hline 
\hline 
$0.020$  & $0.0789151013282$  & $0.0831149115123$  & $0.0831316467916$\tabularnewline
\hline 
$0.040$  & $0.0688806941568$  & $0.0711199370000$  & $0.0711277471472$\tabularnewline
\hline 
$0.050$  & $0.0582435091646$  & $0.0591242044255$  & $0.0591267047363$\tabularnewline
\hline 
$0.055$  & $0.0489342571491$  & $0.0492080726555$  & $0.0492080726555$\tabularnewline
\hline 
$0.058$  & $0.0392459736812$  & $0.0392913669397$  & $0.03929143951989$\tabularnewline
\hline 
$0.059$  & $0.0336811084457$  & $0.0336914844331$  & $0.03369149663655$\tabularnewline
\hline 
$0.0599$  & $0.0221859758365$  & $0.0221860334134$  & $0.02218603343118$\tabularnewline
\hline 
\end{tabular}
\end{table}

\section{Further Physical Quantities}

Overall, physical quantities such as magnetic susceptibility or specific
heat play fundamental roles in thermodynamics, which are essential
for understanding the behavior of the Hubbard model. Therefore, we
explore these physical quantities, focusing particularly on the anomalous
properties they exhibit.

\begin{figure}
\includegraphics[scale=0.52]{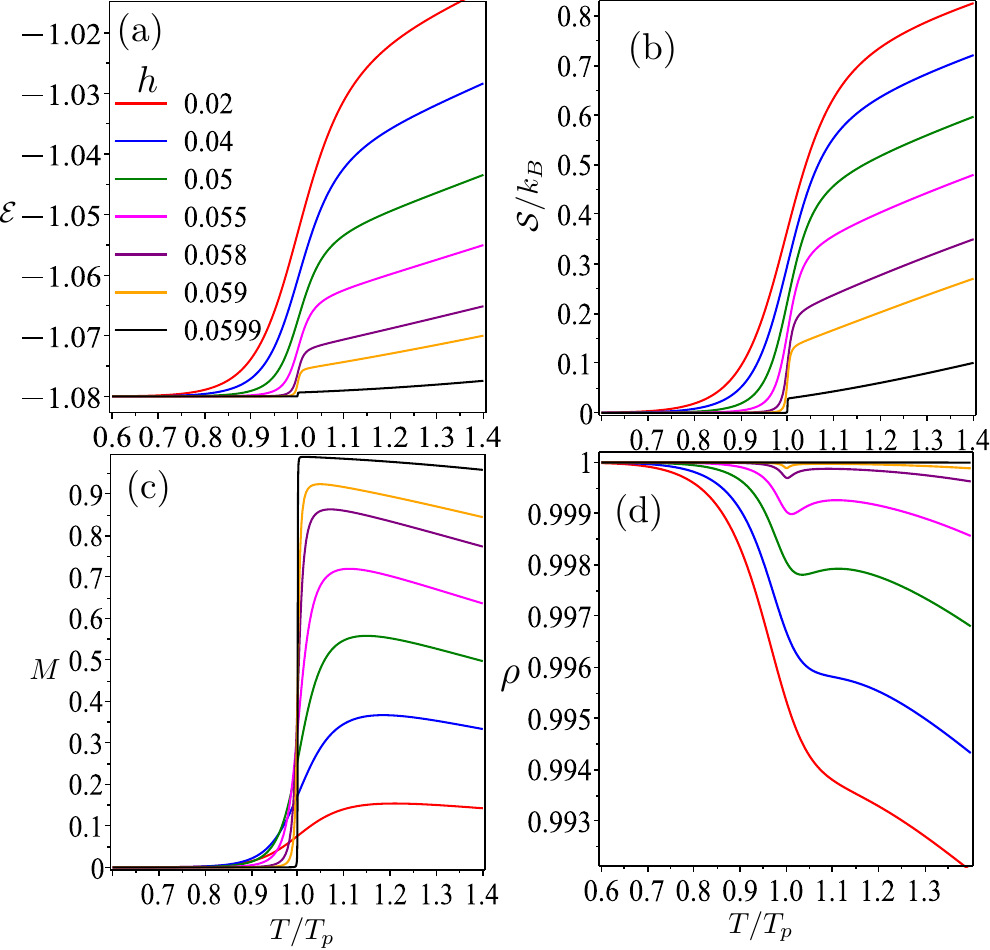} \caption{\protect\label{fig:1fdrv-q} (a) Enthalpy plotted against $T/T_{p}$
for various $T_{p}$ values and magnetic fields as specified in Table
\ref{tab:Psd-crt}, assuming $U=1$ and $V=0.56$. (b) Entropy shown
as a function of $T/T_{p}$. (c) Magnetization represented as a function
of $T/T_{p}$. (d) Electron density depicted as a function of $T/T_{p}$.}
\end{figure}

To provide a clearer depiction of this anomaly, we divide the temperature
$T$ by the peculiar temperature $T_{p}$ listed in Table \ref{tab:Psd-crt},
where the anomaly manifests. Thus in Fig. \ref{fig:1fdrv-q}a, the
enthalpy $\mathcal{E}$ is presented as a function of temperature
normalized by the peculiar temperature ($T/T_{p}$) for various fixed
magnetic fields, as described within the panel. For each fixed magnetic
field, a corresponding peculiar temperature $T_{p}$ is given in Table
\ref{tab:Psd-crt}. At a magnetic field of $h=0.0599$, the enthalpy
$\mathcal{E}$ remains almost constant, but there is a noticeable
``jump'' at $T/T_{p}=1$. This indicates a change from the quasi-alternating
pair $qAP$ region to a one-electron per site region (almost polarized
region) or simply a quasi-paramagnetic $qPM$ region. As the external
magnetic field decreases, this sharp boundary becomes more gradual,
although the ``jump'' in enthalpy becomes larger. Similarly, in
panel (b), the entropy is observed under the same conditions as in
(a), highlighting these features. Notably, for a magnetic field $h\ensuremath{\lesssim}0.06$,
the boundary between $qAP$ and $qPM$ regions becomes more distinct.
As $h$ approaches $0.06$, this characteristic resembles a first-order
or discontinuous phase transition, despite the absence of actual discontinuity.
Panel (c) depicts the magnetization $M$ as a function of $T/T_{p}$.
This allows for the corroboration of the spin orientation in the $qAP$
region, where the magnetization is nearly null, while in the $qPM$
region, the magnetization is almost saturated. This feature is more
pronounced at weaker magnetic fields, becoming smoother as the magnetic
field decreases. Finally, panel (d) illustrates the electron density
$\rho$ as a function of $T/T_{p}$, using the same set of magnetic
fields considered in panel (a). Here, it is observed that the electron
density remains almost constant $\ensuremath{\rho\approx1}$ (but
always smaller than $1$, $\rho<1$) for temperatures $T/T_{p}<1$,
corresponding to a half-filled band. At $T/T_{p}=1$, there is a small
but distinct depression in electron density, which is sharper and
more pronounced at higher magnetic fields. As expected, for lower
magnetic fields, the electron density decreases with increasing temperature.

\begin{figure}
\includegraphics[scale=0.52]{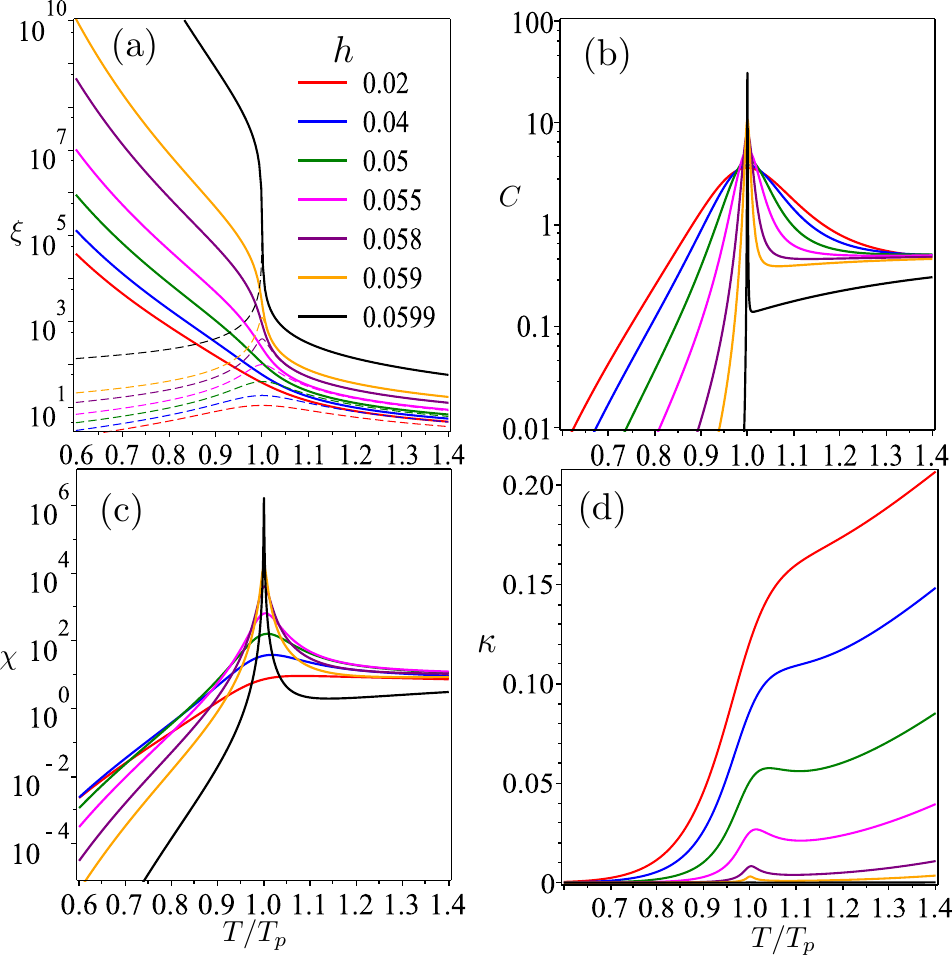} \caption{\protect\label{fig:2fdrv-q} (a) Correlation length plotted against
$T/T_{p}$, with $T_{p\ensuremath{}}$ values provided in Table \ref{tab:Psd-crt},
assuming $U=1$ and $V=0.56$. Dashed lines describe the function
$1/\ln(\lambda_{0}/\lambda_{1})$. (b) Specific heat variation with
$T/T_{p}$. (c) Magnetic susceptibility versus $T/T_{p}$. (d) Isothermal
electron compressibility as a function of $T/T_{p}$.}
\end{figure}

In Fig. \ref{fig:2fdrv-q}a, the correlation length $\xi$ is illustrated
as a function of $T/T_{p}$ for a fixed external magnetic field, as
specified inside the panel. It is important to note that for each
fixed magnetic field, there is a corresponding peculiar temperature
$T_{p}$, as listed in Table \ref{tab:Psd-crt}. The correlation length,
$\xi$ (defined in Sec. IIB, Eq. \eqref{xi}, and represented by solid
lines), demonstrates a decreasing function with an inflection point
around $T/T_{p}=1$. As the magnetic field decreases, the inflection
curvature vanishes. Contrary to what has been previously reported
in the literature \cite{pseudo,unv-cr-exp,psd-hub-dmd,pimenta,BJP20},
$\xi$ does not show the typical peak observed around the pseudo-transition
\cite{pseudo,unv-cr-exp,psd-hub-dmd,pimenta,BJP20}. However, the
function $1/\ln(\lambda_{0}/\lambda_{1})$ (depicted by dashed lines)
exhibits a sharp peak as it approaches $h=0.06$ at $T/T_{p}=1$.
The peak of this function mimics a typical peak of $\xi$ around the
pseudo-transition found in references \cite{pseudo,unv-cr-exp,psd-hub-dmd,pimenta,BJP20},
indicative of a swift change between $qAP$ and $qPM$ regions. However,
as the magnetic field is decreased, this sharp peak becomes more gradual.
Similarly, panel (b) also illustrates the specific heat $C$ as a
function of $T/T_{p}$, using the same set of magnetic fields. Here,
we can observe a sharp peak, akin to that seen in a second-order phase
transition, although there is no actual divergence. Furthermore, panel
(c) displays the magnetic susceptibility plotted against $T/T_{p}$,
clearly showing the sharp peak at $T/T_{p}=1$, which mimics a second-order
phase transition. Finally, panel (d) demonstrates the behavior of
isothermal compressibility $\kappa$. As $h$ approaches $0.06$,
$\kappa$ diminishes, indicating that the system becomes less compressible
under these conditions. Conversely, for smaller magnetic fields, $\kappa$
increases, suggesting that the system is more easily compressible.
At relatively low magnetic fields, $\kappa$ exhibits a minor peak
with a maximum at $T/T_{p}=1$, though this peak becomes less pronounced
with lower magnetic fields.

It is worth mentioning that such a quantity as the entropy always
tends to increase as $T$ grows in accordance with the second law
of thermodynamics, which indicates the thermal stability of the system.
Similarly, the positivity of specific heat and isothermal compressibility
around anomalous behavior also serves as an indicator of stability.

\begin{figure}
\includegraphics[scale=0.5]{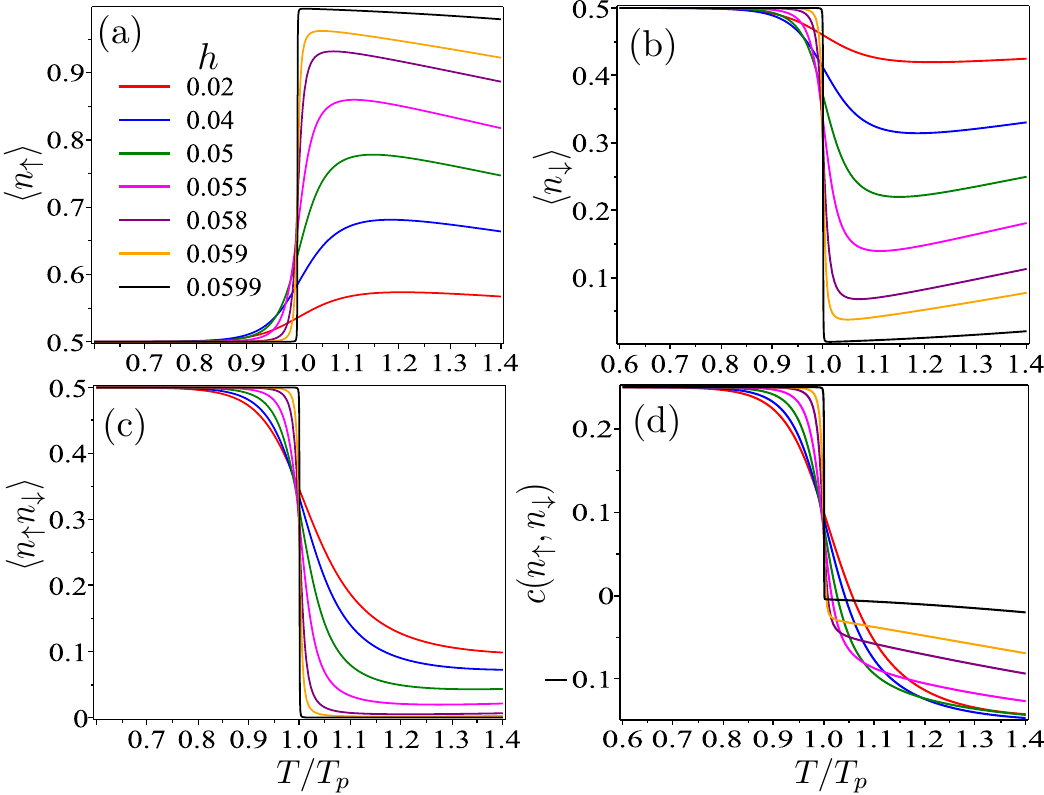} \caption{\protect\label{fig:<n>s} (a) Electron density for spin-up, $\langle\boldsymbol{n}_{\uparrow}\rangle$,
as a function of temperature $T/T_{p}$ across various weak magnetic
fields, assuming $U=1$ and $V=0.56$. (b) Electron density for spin-down,
$\langle\boldsymbol{n}_{\downarrow}\rangle$, as a function of temperature
$T/T_{p}$, corresponding to the magnetic fields used in (a). (c)
On-site average, $\langle\boldsymbol{n}_{\uparrow}\boldsymbol{n}_{\downarrow}\rangle$,
as a function of temperature $T/T_{p}$, assuming the magnetic field
values from (a). (d) On-site correlation function, $c(\boldsymbol{n}_{\uparrow},\boldsymbol{n}_{\downarrow})$,
as a function of temperature $T/T_{p}$, for the same set of parameters
as in (a).}
\end{figure}

In Fig. \ref{fig:<n>s}a, the spin-up electron density $\langle\boldsymbol{n}_{\uparrow}\rangle$
is illustrated as a function of temperature $T/T_{p}$, for several
weak magnetic fields. Notably, in the half-filled band condition,
it is clear how the spin-up averages are organized: for $T/T_{p}<1$,
the spin-up average alternates, given by $\langle\boldsymbol{n}_{\uparrow}\rangle\approx0.5$,
with the remaining spins arranged as spin-down. For $T/T_{p}>1$,
most spins are upwardly arranged. The higher the magnetic field (but
$h<0.6$), the stronger the curvature change at $T/T_{p}=1$. Similarly,
in panel (b), the spin-down electron density $\langle\boldsymbol{n}_{\downarrow}\rangle$
is reported as a function of temperature $T/T_{p}$, for the same
set of magnetic fields as in panel (a). Here, for $T/T_{p}<1$, almost
half of the spins are arranged with spin-down on average, while the
remaining are spin-up, as illustrated in panel (a). In contrast, for
$T/T_{p}>1$, there are nearly no spin-up electrons, as most spins
align with the external magnetic field. Another interesting quantity
is the on-site electron average $\langle\boldsymbol{n}_{\uparrow}\boldsymbol{n}_{\downarrow}\rangle$
as a function of temperature $T/T_{p}$, assuming the same magnetic
field values as in panel (a). This confirms the electron arrangement
in the low-temperature region: For $T/T_{p}<1$, the average leads
to $\langle\boldsymbol{n}_{\uparrow}\boldsymbol{n}_{\downarrow}\rangle\approx0.5$,
indicating that most spins are arranged in pairs alternating with
empty sites. For $T/T_{p}>1$, $\langle\boldsymbol{n}_{\uparrow}\boldsymbol{n}_{\downarrow}\rangle\rightarrow0$,
as most spins are parallel and aligned with the external magnetic
field. For a very weak magnetic field, the curvature change is evident,
and it diminishes as the magnetic field increases. Similarly, in panel
(d), the correlation function $c(\boldsymbol{n}_{\uparrow},\boldsymbol{n}_{\downarrow})=\langle\boldsymbol{n}_{\uparrow}\boldsymbol{n}_{\downarrow}\rangle-\langle\boldsymbol{n}_{\uparrow}\rangle\langle\boldsymbol{n}_{\downarrow}\rangle$
is depicted as a function of temperature $T/T_{p}$, again for the
same set of parameters as in panel (a). For $T/T_{p}<1$, the correlation
function is roughly $c(\boldsymbol{n}_{\uparrow},\boldsymbol{n}_{\downarrow})=0.25$,
while for $T/T_{p}>1$, it becomes negative, as $\langle\boldsymbol{n}_{\uparrow}\boldsymbol{n}_{\downarrow}\rangle\approx0$,
$\langle\boldsymbol{n}_{\uparrow}\rangle\approx0.5$ and $\langle\boldsymbol{n}_{\downarrow}\rangle\approx0.5$,
leading to a negative correlation function of approximately $\approx-0.25$.

\section{Conclusion }

Our study has provided a comprehensive re-examination of the one-dimensional
extended Hubbard model in the atomic limit, a subject initially explored
in the 1970s. This result not only confirms previous outcomes of the
model but also reveals novel insights, especially the occurrence of
finite-temperature pseudo-transitions in specific quasi-one-dimensional
models.

Initially, our analysis focused on delineating the zero-temperature
phase diagram at zero magnetic field. Here, we identified three distinct
types of frustrated phases, enriching our understanding of the intricate
phase structure within the model. This phase characterization was
facilitated by the employment of the transfer matrix technique, followed
by precise numerical analyses, which unveiled anomalous behaviors
within the half-filled band of the model. These behaviors are particularly
pronounced in the low-temperature regime, especially during the gradual
shift between alternating pair ($AP$) and paramagnetic ($PM$) phases.
The subtlety of these anomalous behaviors, possibly overlooked in
earlier studies due to the deceptively simple numerical facade of
the model, but from an algebraic perspective, requiring the solution
of an algebraic cubic equation introduces a level of slightly more
elaborate manipulation, revealing the intricate nature of the model.

Further investigations explored the low-temperature pseudo-phase transitions
at the half-filling band, where the aforementioned anomalous behavior
is predominantly observed. We observed that close to this anomalous
region, a pseudo-transition emerges that exhibits characteristics
reminiscent of both first- and second-order phase transitions. This
pseudo-transition is particularly fascinating, as it mimics the intricacies
of real phase transitions, adding a layer of sophistication to our
understanding of the model. Moreover, our exploration extended to
diverse properties such as electron density, on-site correlations,
and nearest-site electron averages. In examining these quantities,
we discerned the underlying reasons for the occurrence of pseudo-transitions,
thereby deepening our grasp of the model behavior. The intrinsic nature
of the model, defined by its nearest-neighbor interaction, precludes
the occurrence of real phase transitions at finite temperatures, thereby
accentuating the distinctive properties and behaviors inherent to
the model within a low-dimensional system and underscoring its peculiarity.

The property we have explored may not necessarily be exclusive to
one-dimensional models; however, uncovering this anomalous residual
boundary entropy could become a challenging task in higher dimensional
systems.
\begin{acknowledgments}
O.~R. and S.~M.~S. thank CNPq and FAPEMIG for partial financial
support. M.~L.~L. and M.~S.~S.~Pereira also acknowledge the financial
support of FAPEAL. O.~D. acknowledges support through the EURIZON
project (project \#3025), which is funded by the European Union under
grant agreement No.871072. L.M.V acknowledges the support of Donostia
International Physics Center and iKur estrategia.
\end{acknowledgments}

\appendix

\section{Transfer matrix around pseudo-transition \protect\label{sec:Appdx=000020A}}

The phenomenon of pseudo-transition arises near the alternating pairs
$AP$ and paramagnetic $PM$ phases. Although the solution of \eqref{eq:sol-cub}
is exact and analytic, the analytical expression for finding the peculiar
temperature could becomes a cumbersome task due to involving cubic
root expression. An alternative approach to determine this condition
is to split the transfer matrix into two terms, namely, $\mathbf{V}=\mathbf{V}_{0}+\varsigma\mathbf{V}_{1}$.
Here, $\mathbf{V}_{0}$ represents the core structure matrix, whose
matrix elements just includes the half-filled case, while $\varsigma\mathbf{V}_{1}$
describes a small perturbation applied to the matrix, corresponding
to non-half-filled band. The detailed formulations of these matrices
are provided as follows: 
\begin{equation}
\mathbf{V}_{0}=\left[\begin{array}{ccc}
0 & 0 & w_{0,2}\\
0 & w_{1,1}z & 0\\
w_{0,2} & 0 & 0
\end{array}\right]
\end{equation}
and 
\begin{equation}
\mathbf{V}_{1}=\left[\begin{array}{ccc}
1 & w_{0,1}\sqrt{z} & 0\\
w_{0,1}\sqrt{z} & 0 & w_{1,2}\sqrt{z}\\
0 & w_{1,2}\sqrt{z} & w_{2,2}
\end{array}\right],
\end{equation}
respectively. In this context, $\varsigma$ acts as a small formal
parameter that, for our purposes, is set to 1 to measure how the system
deviates from the half-filled-band limit. This approach allows us
to analyze the effects of slight perturbations on the system by examining
the changes introduced by $\mathbf{V}_{1}$ in relation to the original
matrix $\mathbf{V}_{0}$.

Therefore, diagonalizing $\mathbf{V}_{0}$, we have the following
eigenvalues 
\begin{alignat}{1}
v_{0}^{(0)}= & w_{1,1}z,\label{V011}\\
v_{1}^{(0)}= & w_{0,2},\label{V002}\\
v_{2}^{(0)}= & -w_{0,2}.
\end{alignat}
From this solution one can clearly verify that $v_{1}^{(0)}>v_{2}^{(0)}$,
however, we cannot affirm any condition between $v_{0}^{(0)}$ and
$v_{1}^{(0)}$, since there is no restriction, so one can establish
the following condition $v_{1}^{(0)}=v_{0}^{(0)}$ this would be essential
to find the peculiar temperature, as discussed in Sec. \ref{subsec:Psd-crt}.
The corresponding eigenvectors can be expressed by 
\begin{alignat}{1}
|u_{0}^{(0)}\rangle= & |\updownarrow\rangle,\\
|u_{1}^{(0)}\rangle= & \tfrac{1}{\sqrt{2}}\left(|0\rangle+|\updwn\rangle\right),\\
|u_{2}^{(0)}\rangle= & \tfrac{1}{\sqrt{2}}\left(-|0\rangle+|\updwn\rangle\right).
\end{alignat}

Now let us improve our previous result, thus the first-order correction
on the transfer matrix eigenvalues can be obtained perturbatively,
resulting in 
\begin{alignat}{1}
v_{0}^{(1)}= & 0,\label{eq:V1_11}\\
v_{1}^{(1)}= & \tfrac{1}{2}\left(1+w_{2,2}\right),\label{eq:V1_02}\\
v_{2}^{(1)}= & \tfrac{1}{2}\left(1+w_{2,2}\right).
\end{alignat}
Here evidently we have the following condition $v_{0}^{(1)}<v_{1}^{(1)}=v_{2}^{(1)}$,
because $w_{2,2}>0$.

Similarly, the second-order corrections of eigenvalues becomes, 
\begin{alignat}{1}
v_{0}^{(2)}= & \tfrac{z\left(w_{0,1}+w_{1,2}\right)^{2}}{2\left[w_{1,1}z-w_{0,2}\right]}+\tfrac{z\left(w_{0,1}-w_{1,2}\right)^{2}}{2\left[w_{1,1}z+w_{0,2}\right]},\\
v_{1}^{(2)}= & \tfrac{1}{8}\tfrac{\left(w_{2,2}-1\right)^{2}}{w_{0,2}}-\tfrac{z\left(w_{0,1}+w_{1,2}\right)^{2}}{2\left[w_{1,1}z-w_{0,2}\right]},\\
v_{2}^{(2)}= & -\tfrac{1}{8}\tfrac{\left(w_{2,2}-1\right)^{2}}{w_{0,2}}-\tfrac{z\left(w_{0,1}-w_{1,2}\right)^{2}}{2\left[w_{1,1}z+w_{0,2}\right]}.
\end{alignat}
It is evident that $u_{1}^{(2)}>u_{2}^{(2)}$ and $u_{2}^{(2)}<0$.

Therefore, the corresponding eigenvalues up to order ${\cal O}(\varsigma^{3})$,
are given approximately as follow 
\begin{alignat}{1}
\lambda_{j}= & v_{j}^{(0)}+\varsigma v_{j}^{(1)}+\varsigma^{2}v_{j}^{(2)}+{\cal O}(\varsigma^{3}).
\end{alignat}
It is evident that the eigenvalues up to second-order correction indicate
that $\lambda_{1}>\lambda_{2}$. However, we cannot affirm just by
looking the perturbative eigenvalues which eigenvalues is the largest
one $\lambda_{0}$ or $\lambda_{1}$. Although we have confirmed at
the end of Sec. \ref{subsec:Transfer-matrix} that $\lambda_{0}$
must be the largest one. Nevertheless, our perturbative result could
be useful to find the condition of peculiar temperature up to second-order
approximation. Therefore, by using the condition given in \eqref{eq:g-cond}
we can find the following relation 
\begin{equation}
w_{1,1}z-w_{0,2}+\tfrac{3\left(1+w_{2,2}\right)}{2}-\tfrac{1}{8}\tfrac{\left(w_{2,2}-1\right)^{2}}{w_{0,2}}-\tfrac{z\left(w_{0,1}-w_{1,2}\right)^{2}}{2\left[w_{1,1}z+w_{0,2}\right]}=0,
\end{equation}
this result would be relevant when we look for a more accurate value
of the peculiar temperature.

\end{document}